%% file: skeleton.tex
\documentclass[a4paper,11pt]{article}
\usepackage{pos,wrapfig,enumitem,xspace}

\newcommand{\eV}{\mathrm{eV}}
\newcommand{\EeV}{\mathrm{EeV}}
\newcommand{\km}{\mathrm{km}}
\newcommand{\yr}{\mathrm{yr}}
\newcommand{\sr}{\mathrm{sr}}
\providecommand{\degree}{^\circ}
\newcommand{\lgE}{\log_{10}(E/\mathrm{eV})}

\newcommand{\dd}{\operatorname{d}\!}
\usepackage{multirow,multicol,lineno}

\title{The UHECR dipole and quadrupole in the latest data from the original Auger and TA surface detectors}
\ShortTitle{The UHECR dipole and quadrupole}
\author*[a]{Peter Tinyakov}
\author{Luis Anchordoqui}\author{Teresa Bister}\author{Jonathan Biteau}\author{Lorenzo~Caccianiga}\author{Rogério de~Almeida}\author{Olivier Deligny}\author{Armando di~Matteo}\author{Ugo~Giaccari}\author{Diego Harari}\author{Jihyun Kim}\author{Mikhail Kuznetsov}\author{Ioana Mariș}\author{Grigory~Rubtsov}\author{Sergey Troitsky}\author{Federico Urban}
\affiliation[a]{Service de Physique Théorique, Université Libre de Bruxelles,\\ Boulevard de la Plaine, 1050~Brussels, Belgium}
\affiliation[b]{Observatorio Pierre Auger, Av.\ San Mart{\'\i}n Norte 304, 5613 Malarg\"ue, Argentina
}
\affiliation[c]{Telescope Array Project, 201 James Fletcher Bldg, 115 S.\ 1400 East, Salt Lake City, UT 84112-0830, USA
}
\forColl{Pierre~Auger$^b$}\forColl{ and the Telescope Array$^c$}
\emailAdd{spokespersons@auger.org}
\emailAdd{ta-icrc@cosmic.utah.edu}

\abstract{The sources of ultra-high-energy cosmic rays are still unknown, but assuming standard physics, they are expected to lie within a few hundred megaparsecs from us.  Indeed, over cosmological distances cosmic rays lose energy to interactions with background photons, at a rate depending on their mass number and energy and properties of photonuclear interactions and photon backgrounds. The universe is not homogeneous at such scales, hence the distribution of the arrival directions of cosmic rays is expected to reflect the inhomogeneities in the distribution of galaxies; the shorter the energy loss lengths, the stronger the expected anisotropies.  Galactic and intergalactic magnetic fields can blur and distort the picture, but the magnitudes of the largest-scale anisotropies, namely the dipole and quadrupole moments, are the most robust to their effects.  Measuring them with no bias regardless of any higher-order multipoles is not possible except with full-sky coverage. In this work, we achieve this in three energy ranges (approximately $8$--$16~\EeV$, $16$--$32~\EeV$, and $32$--$\infty~\EeV$) by combining surface-detector data collected at the Pierre Auger Observatory until 2020 and at the Telescope Array (TA) until 2019, before the completion of the upgrades of the arrays with new scintillator detectors.  We find that the full-sky coverage achieved by combining Auger and TA data reduces the uncertainties on the north-south components of the dipole and quadrupole in half compared to Auger-only results.}

\FullConference{37$^{\rm{th}}$ International Cosmic Ray Conference (ICRC 2021)\\
		July 12th -- 23rd, 2021\\
		Online -- Berlin, Germany}

\begin{document}
\maketitle

\section{Introduction}

While on general grounds the anisotropies of cosmic rays (CR) at highest
energies are expected to give a key to understanding of their sources, in
practice the deflections of cosmic rays in (highly uncertain) magnetic fields
make this task extremely difficult. Even though the angular resolution in the 
reconstruction of CR arrival directions is good (${\sim}1^\circ$), no anisotropies 
have been discovered at small angular scales of order a few degrees, likely due to a washing effect of the
magnetic deflections. 

Even when individual sources are not resolved, at large angular scales the
anisotropies are expected to arise from a non-homogeneous source distribution
in the Universe at distances of 50--100~Mpc. The effect of magnetic deflections
at these angular scales, particularly for the dipole and quadrupole harmonics,
may not be dominant. The non-zero dipole anisotropy in the CR
distribution, with an amplitude of $6.5\%$ in the equatorial plane, has
indeed been discovered by the Pierre Auger Observatory (hereafter Auger) \cite{bib:AugerScience2017} at energies
$E>8$~EeV.  
The equatorial dipole has also been
measured by the Telescope Array (TA) \cite{Abbasi:2020ohd}, with the result in agreement with
that of Auger but not significant alone due to small statistics.
 
Both Auger and TA observatories have incomplete sky coverage, which makes the
unambiguous determination of all multipole components impossible.  To achieve the
full-sky coverage, they have to be combined together.  However, this cannot be
trivially done by simply combining the events detected by the two observatories
with their nominal energy estimates, as they both have
potentially different systematic uncertainties in the energy determination, and these
differences may affect the sky distribution of events in a given energy range
in the combined data set. The energy scales have to be cross-calibrated before
combining the data. This can be done from the data themselves without any
extra assumptions on the nature of the energy determination systematics
\cite{bib:ApJ2014} by comparing the data of the two observatories in the
equatorial band where their exposures overlap.

The purpose of this study is to cross-calibrate energy scales of the
Auger and TA observatories in the energy range $E\gtrsim 10$~EeV and to
determine in an assumption-free way the dipole and quadrupole 
components of the UHECR flux in this energy range. This study is 
a development of the previous analyses \cite{bib:ApJ2014,bib:UHECR2016,bib:UHECR2018} in the
following three respects:  (i) we use the updated data sets, (ii) we
cross-calibrate the energy scales in three energy bins relaxing the
assumption that the difference in energy scales is energy-independent and
(iii) we carefully trace all sources of uncertainties and propagate them into the
final result. 

\section{The datasets}
The Pierre Auger Observatory \cite{bib:Auger}
is a hybrid detector of UHECRs 
located in the Southern hemisphere in Argentina at a latitude of $-35.2^\circ$. It consists of a surface array of 1660 water-Cherenkov detectors covering an area of approximately 3000~km$^2$, overlooked by the fluorescence detector composed of 27 fluorescence telescopes. The detector is taking data since January 2004. In this work, we use the dataset described in Ref.~\cite{bib:AugerLSA2021}, consisting of events detected by the surface detector (SD) array from 2004 Jan~01 to 2020 Dec~31.  Its geometric exposure
is~$110\,000~\km^2~\yr~\sr$.
In order to correct for the effects of the finite energy resolution, which in the case of a decreasing energy spectrum tend to make the raw measured spectrum higher than the actual one, we divide the geometric exposures by the unfolding correction factors reported in Ref.~\cite{bib:AugerPRD2020}, which in the energy range used in this work, increase from~0.977 in the energy bin~$18.9 \le \lgE < 19.0$ to~1.002 in the bin~$19.6 \le \lgE < 19.7$ then decrease again to~0.964 in the bins~$\lgE \ge 19.8$.  
The effects of the tilt of the SD array and of the non-uniformity of its aperture in sidereal time were found to be minor in Ref.~\cite{bib:AugerScience2017} (of the order of $1/4$~and $1/20$~of the statistical uncertainties, respectively) and are neglected in this work.

The Telescope Array \cite{TAdetector} 
is a hybrid detector of UHECRs located in the Northern hemisphere in Utah, USA at a latitude of $39.3^\circ$. It is taking data since May 2008. The surface detector of TA consists of 507 plastic scintillator detectors covering an area of about 700~km$^2$. The fluorescence detector of TA is composed of 38 fluorescence telescopes arranged in 3 towers overlooking the surface detector area.
In this work, we use the events with the zenith angles $\theta<55^\circ$ detected by the TA SD array from 2008 May~11 to 2019 May~10 with the selection criteria as described in \cite{bib:ApJ2014}.  The effective exposure, accounting for the effects of the energy resolution, in the energy range used in this work depends non-monotonically on energy, with a minimum value of~$13\,200~\km^2~\yr~\sr$ in the energy bin~$19.7 \le \lgE < 19.8$, a maximum of~$15\,400~\km^2~\yr~\sr$ in~$20.0 \le \lgE < 20.1$, and a value of~$15\,100~\km^2~\yr~\sr$ in the bins~$\lgE \ge 20.2$.

\newcommand{\n}{\mathbf{n}}
\newcommand{\PA}{\text{Auger}}
\newcommand{\TA}{\text{TA}}
\newcommand{\effA}{\mathcal{E}}
\newcommand{\og}{\omega_\text{geom}}
We assume that the effective exposure of each detector array 
factorizes into an effective area~$\effA$ which depends on the energy~$E$ but not on the arrival direction, and a geometric exposure~$\og$ which only depends on the declination as described in~Ref.~\cite{bib:exposure}.
In the case of Auger, we consider two different effective areas for vertical~($\theta < 60\degree$) and inclined~($\theta \ge 60\degree$) events due to the different reconstruction techniques used for these two zenith angle ranges.
\begin{figure}
    \centering
    \includegraphics{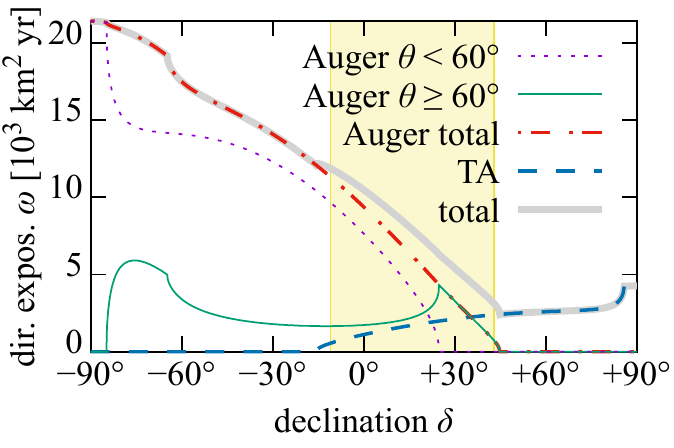}
    \includegraphics{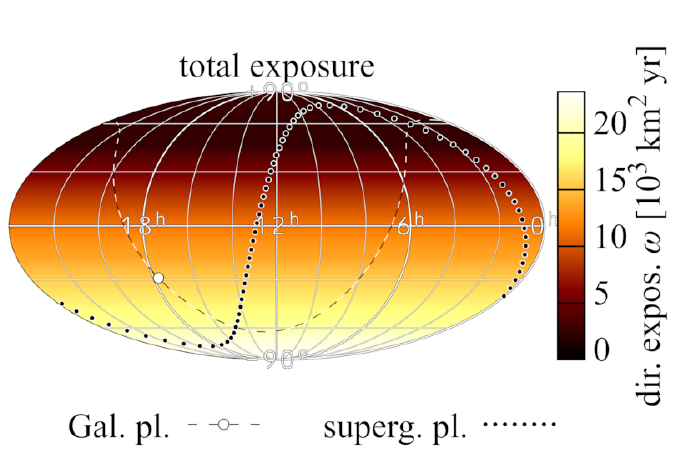}
    \caption{The combined Auger + TA effective exposure in the first energy bin, see Sect.~\ref{sec:crosscalibration}. The yellow band on the left panel indicates the range of declinations visible to both observatories used for the cross-calibration of energy scales. }
    \label{fig:exposure}
\end{figure}
As an example, the combined exposure averaged over the lowest of the three energy bins in which we search for anisotropies (see Sect.~\ref{sec:crosscalibration}) is shown in Fig.~\ref{fig:exposure}.


\section{The cross-calibration of energy scale}
\label{sec:crosscalibration}

As already mentioned, when combining data from different detectors to infer the large-scale anisotropy, it is important to ensure that the energy thresholds used for them match, otherwise spurious detections of the north-south anisotropy may arise. 
To eliminate this problem, we calibrate the Auger and TA energy scales to each other using the events detected in the equatorial region of the sky where both observatories have exposure (see Fig.~\ref{fig:exposure}). 
The idea behind the cross-calibration procedure is that the true UHECR flux integrated over the common band as a function of energy, $\phi^{\rm true}(E)$, can be estimated in each of the two observatories independently. When the energy scales match, these two estimates must agree at all energies. Note that this procedure does not allow one to determine the true energy scale, only to eliminate a possible mismatch in scales between the two observatories. This is sufficient, however, for the purpose of the present analysis. 

To estimate the flux we use the unbiased estimator introduced in \cite{bib:spectrum2016}, which is essentially the sum over events in the band weighted with the inverse exposure $\omega_a$ of each observatory
$a=(\text{Auger, TA})$. In the energy bin $j$, defined in terms of the 
nominal energy of the corresponding observatory,
the flux estimate is
    $\phi^\text{est.}_{aj} = \sum_k 
    1/\omega_{a}(\n_k)$,
where $\n_k$ is the arrival direction of a given event and the sum runs over the events $k$ with energies $E_{ak} \in [E_{aj}, E_{aj+1})$ and arrival directions in the common band $\delta_k \in (\delta_{\min}, \delta_{\max})$. 
It can be shown that, provided the  exposure does not vanish anywhere in the band and regardless of the variations of the flux density $J(E,\n)$ over the sky, the quantity~$\phi^\text{est.}_{aj}$ is an unbiased estimator of~$\phi^\text{true}_{j} = \int_{E_{aj}}^{E_{aj+1}} \int_{\delta_{\min}}^{\delta_{\max}} J(E,\n)\,\mathrm{d}E\,\mathrm{d}\Omega$.
The latter is by definition a detector-independent quantity, and hence,
in the absence of energy scale differences, the two estimators $\phi^\text{est.}_{{\rm Auger,}j}$ and $\phi^\text{est.}_{{\rm TA,}j}$ must agree within statistical uncertainties in all energy bins. The cross-calibration consists in determining the energy conversion function between the two observatories such that this requirement is satisfied. Unlike in Refs.~\cite{bib:UHECR2016,bib:UHECR2018}, where the cross-calibration was applied to energy thresholds --- which is equivalent to assuming a constant energy rescaling factor --- in this work we assume a functional form for the $E_\text{TA} \leftrightarrow E_\text{Auger}$~conversion with free parameters which we fit to satisfy the flux matching conditions in the common band. 

The fiducial boundaries of the common band are chosen as follows. The intersection of the Auger and TA fields of view is $(-15.7\degree, +44.8\degree)$, but in the areas close to the edge $1/\omega_i$~becomes very large, which would result in increased statistical uncertainties if the entire intersection was used.  Instead, we choose a fiducial band~$(\delta_{\min} = -11\degree, \delta_{\max} = +43\degree)$, which are the values that minimize the expected total statistical uncertainties 
rounded to the nearest degree.  

We  bin the events in logarithmic nominal energy bins defined as $\log_{10}(E_{\PA, j}/\eV) = 18.9 + 0.1 j$ and~$\log_{10}(E_{\TA, j}/\eV) = 19.0 + 0.1 j$, for~$j = 0$, $1$, .\,.\,\@. The last non-empty bin is~$[20.1, 20.2)$ for~Auger and~$[20.2, 20.3)$ for~TA. We then compute the corresponding statistical uncertainties of the flux estimators as $\sigma_{aj} = \sqrt{\sum_k 1/ \omega_{a}(E_k, \n_k)^2 }$, where the sum runs over the same events as above.

\newcommand{\Etrue}{E}
\newcommand{\Ecorr}{E_\text{ref}}
\newcommand{\params}{{\boldsymbol{\theta}}}

\newcommand{\lnshift}{\alpha}
\newcommand{\exponent}{\beta}
\newcommand{\gA}{{\gamma_\mathrm{A}}}
\newcommand{\EAB}{{E_\mathrm{AB}}}
\newcommand{\gB}{{\gamma_\mathrm{B}}}
\newcommand{\EBC}{{E_\mathrm{BC}}}
\newcommand{\gC}{{\gamma_\mathrm{C}}}
We assume a power law for the energy conversion in the range $E_\TA \ge 10~\EeV$ characterized by two parameters
$\params_E = (\lnshift,\exponent)$:
\begin{align}
    E_\PA & = E_0 e^\alpha  \left(E_\TA/E_0\right)^\beta ,\nonumber \\
    E_\TA & = E_0 e^{-\alpha/\beta}  \left(E_\PA/E_0\right)^{1/\beta}
    \label{eq:conv}
\end{align} 
with $E_0=10~\EeV$. 
The goal of the cross-calibration is to determine the parameters $\alpha$ and $\beta$. If we binned the events in corrected energies, events would move from one bin to another when the parameters are changed, producing discontinuities in the flux estimator. To avoid this technical problem, we fit both Auger and TA data to a model spectrum in the band while keeping both sets of nominal energy bins fixed. For the model spectrum we use the twice-broken\footnote{
    We find that using instead the smoothed breaks as in Ref.~\cite{bib:AugerPRD2020} would make only a minor difference in the goodness of fit ($\Delta \chi^2 = -0.3$, $\Delta p = 0.02$) and a negligible difference in the energy conversion ($\Delta\lnshift = 4\times10^{-4}$, $\Delta\exponent = -10^{-3}$).
} power law ($\propto E^{-\gamma}$) with normalization~$\mathcal{J}_0$, break energies~$\EAB$ and~$\EBC$ and spectral indices~$\gA$, $\gB$ and~$\gC$, for a total of six additional parameters~$\params_J$.
For each set of resulting 8 parameters, we compute the model predictions
\(
    \phi^\text{pred}_{j}(\params_J,\params_E) = 
    \int_{E_{j}}^{E_{j+1}}
    \mathcal{J}_\text{band}(E; \params_{J})\dd{\Etrue}
\)
where $E$ is some (arbitrarily chosen) energy scale and the bin boundaries $E_i$ are obtained by converting the bins of the corresponding observatory into that scale. 
Finally, we compute $\chi^2$ according to a log-normal distribution, \begin{equation}
    \chi^2 = \sum_{aj}
    \frac
    {\bigl(\ln\phi_{aj}^\text{est.} - \ln\phi^\text{pred.}_{aj}(\params_J,\params_E) \bigr)^2}
    {\bigl(\sigma_{aj}\big/\phi^\text{est.}_{aj}\bigr)^2},
\end{equation}
which we found to adequately describe the probability distribution of the flux estimator in simulations provided there are ${\gtrsim}10$~events in the fiducial band in each energy bin, which we achieve by combining together the last energy bins of each dataset until this condition is achieved (for~Auger, 19~events with~$\lgE \ge 19.9$; for~TA, 11~events with~$\lgE \ge 20.0$, remaining with 11~bins for both Auger and TA). 
\begin{table}
    \centering
    \begin{tabular}{cc|r|cccccc|}
    \cline{3-9}
    ~ & ~ & ~ &
    $\frac{\mathcal{J}_0}{\eV^{-1}~\km^{-2}~\yr^{-1}}$ &
    $\gA$ &
    $\frac{\EAB}{\EeV}$ &
    $\gB$ &
    $\frac{\EBC}{\EeV}$ &
    $\gC$ \\
    \cline{3-9}
    $\lnshift=-0.154\pm0.013$ & \quad & 
    Auger scale & $2.65\times10^{-19}$ & 2.53 & 11.7 & 2.92 & 49.6 & 5.66 \\
    $\exponent=\phantom{+}0.937 \pm 0.017$ & \quad &
    TA scale & $3.14\times10^{-19}$ & 2.43 & 13.9 & 2.80 & 65.1 & 5.37 \\
    \cline{3-9}
    $\rho_{\alpha\beta} = -0.177$ & \multicolumn{2}{c}{} &
    \multicolumn{6}{c}{$\chi^2/n_\text{dof} = 15.6/14$ ($p=0.34$)} \\
    \end{tabular}
    \caption{Best-fit parameter values from the spectrum fit used for the cross-calibration procedure.  
    }
    \label{fig:spectrum}
\end{table}
The resulting best fit is shown in \autoref{fig:spectrum}.
If the exponent~$\exponent$ is fixed to~$1$ as corresponds to a constant rescaling factor between Auger and TA energies, the best fit becomes~$\lnshift=-0.163\pm0.012$ with a~$\chi^2/n$ of~$29.1/13$. The energy-dependent rescaling of \autoref{fig:spectrum} is thus favored over a constant one
at the $3.7\sigma$~level.  The possible origins of such an 
energy-dependence are currently being investigated by the two collaborations \cite{bib:spectrumWG}. 

Based on the cross-calibration results, the following energy bin boundaries have been chosen for the calculation of the dipole and quadrupole components:
\begin{center}\begin{tabular}{|c|c|c|c|}
    \hline
    Auger scale & $8.57\pm0.11~\EeV$ & $16~\EeV$ & $32~\EeV$ \\
    \hline
    TA scale & $10~\EeV$ & $19.47\pm0.32~\EeV$ & $40.8\pm1.1~\EeV$ \\
    \hline
\end{tabular}\end{center}
The lowest energy of 10~EeV (TA) is fixed by the availability of the TA events, while the other two boundaries 16~EeV (Auger) and 32~EeV (Auger) are chosen so as to match the previous Auger dipole analysis. The exposures 
in these three bins, appropriately corrected for a slight energy dependence of exposure in each observatory, 
are 
$(87\,400+24\,600)~\km^2~\yr~\sr$, $(87\,200+24\,600)~\km^2~\yr~\sr$ and $(86\,600+24\,400)~\km^2~\yr~\sr$ for Auger vertical + inclined
and $14\,200~\km^2~\yr~\sr$, $14\,000~\km^2~\yr~\sr$ and $13\,700~\km^2~\yr~\sr$ for TA.

The exposures of the two detectors are only known with an uncertainty of approximately~$\pm 3\%$ each.  An over- or under-estimate of the TA-to-Auger exposure ratio by~$\pm 4.2\%$ would result in an over- or under-estimate of the parameter~$\lnshift$ by~$0.023$, i.e.\ of~$E_\text{Auger}(E_\text{TA}=10~\EeV)$ by~$0.20~\EeV$, as well as of~$\exponent$ by~$\mp0.06$, meaning the relative uncertainty shrinks at higher energies.  On the other hand, the effect of such an uncertainty on the final anisotropy results would almost completely cancel out those of the uncertainty in the exposure ratio itself.  For this reason, we neglect this source of uncertainty in the following.

\subsection{Propagation of the statistical uncertainty on the calibration fit}\label{sec:uncertainty}

The uncertainty on the correspondence between energy thresholds causes as an ``effective'' uncertainty on the exposure ratio.
If we overestimate e.g.~the Auger threshold corresponding to a given TA threshold, we underestimate the flux in the Auger field of view. The effect on large-scale anisotropy searches can be approximated to that of overestimating its exposure, provided the anisotropies themselves do not appreciably change with energy over such a range.
To quantify this uncertainty, we propagate the statistical uncertainties of the fit on~$\alpha$ and~$\beta$ (\autoref{fig:spectrum}) to the quantity
\begin{equation} \ln\left(\frac{\phi^\text{pred}_{\TA,j'}(\params_J,\alpha_\text{best},\beta_\text{best})}{\phi^\text{pred}_{\TA,j'}(\params_J,\alpha,\beta)} \frac{\phi^\text{pred}_{\PA,j'}(\params_J,\alpha,\beta)}{\phi^\text{pred}_{\PA,j'}(\params_J,\alpha_\text{best},\beta_\text{best})} \right)\end{equation}
where $\phi^\text{pred}_{ij'}$~is computed as above, but over the ``wide'' energy bins~$j'$ we are going to use for anisotropy searches, not the ``narrow'' energy bins~$j$ used for the cross-calibration
---~in other words, how much we would overestimate the TA-to-Auger flux ratio if the true conversion parameters were $(\alpha,\beta)$ but we assumed they were $(\alpha_\text{best},\beta_\text{best})$.
The result is~$\pm 2.5\%$, $\pm 2.5\%$ and~$\pm 6.5\%$ in the first, second and third energy bin, respectively.  It can be noted that these values are roughly of the order of $(\gamma-1)$~times the relative uncertainties on the energy thresholds, as they would be in the case of a single threshold for an integral flux $\int_E^{+\infty} \mathcal{J}_\text{band}(E) \dd{E} \propto E^{-(\gamma-1)}$.

\section{Results on large-scale anisotropies}
The UHECR flux $\Phi(\n)$ can be represented as a  sum of spherical harmonics $Y_{lm}$, 
\newcommand{\sumlm}{\sum_{\ell=0}^{+\infty}\sum_{m=-\ell}^{+\ell}}
\begin{align}
    \Phi(\n) &= \sumlm a_{\ell m} Y_{\ell m}(\n), &
    a_{\ell m} &= \int_{4\pi} Y_{\ell m}(\n) \Phi(\n) \dd\Omega.
\end{align}
The coefficients $a_{\ell m}$ represent anisotropies on scales $\mathcal{O}(180\degree/\ell)$. The contribution of the two lowest non-trivial harmonics $l=1,2$, the dipole and quadrupole, can be rewritten in terms of a dipole vector $\mathbf{d}$ and the symmetric traceless quadrupole tensor $Q_{ij}$ as follows,
\begin{equation}
    \Phi(\n) = \Phi_0\left( 1 + \mathbf{d}\cdot \n + \tfrac{1}{2}\n \cdot \mathsf{Q}\n + \cdots \right)
\end{equation}
where $\Phi_0 = \sqrt{4\pi}\,a_{00}$, $d_x = \sqrt{3}a_{11}/a_{00}$, $d_y = \sqrt{3}a_{1-1}/a_{00}$, $d_z = \sqrt{3}a_{10}/a_{00}$,
$Q_{xx}-Q_{yy} = 2\sqrt{15} a_{22}/a_{00}$, $Q_{xz} = \sqrt{15} a_{21}/a_{00}$, $Q_{yz} = \sqrt{15} a_{2-1}/a_{00}$, 
$Q_{zz} = 2\sqrt{5} a_{20}/a_{00}$, $Q_{xy} = \sqrt{15} a_{2-2}/a_{00}$  
and the other components of $Q_{ij}$ can be computed from its symmetry and zero trace condition.

Using a full-sky dataset with the combined exposure $\omega(\n)$ non-zero everywhere,
each~$a_{\ell m}$ can be estimated independently as $\sum_k Y_{\ell m}(\n_k)/\omega(\n_k)$.
We estimate the statistic uncertainties and the correlations between them as $\sigma_{\ell m}\sigma_{\ell' m'}\rho_{\ell m,\ell' m'} = \sum_k Y_{\ell m}(\n_k)Y_{\ell' m'}(\n_k)/\omega(\n_k)^2$.
As for the uncertainties due to the energy cross-calibration,
we compute them by computing~$a_{\ell m}^\pm$ assuming the exposure~$\omega^\pm = \omega_\PA + e^{\pm k}\omega_\TA$ where $k$~is the ``effective'' exposure ratio uncertainty described in \autoref{sec:uncertainty}, and taking $\sigma_\text{syst}(d_z) = \frac{1}{2}(d_z^+ - d_z^-)$, $\sigma_\text{syst}(Q_{zz}) = \frac{1}{2}(Q_{zz}^+ - Q_{zz}^-)$, and similarly for other components. 
The rotationally invariant quantities~$C_1 = \tfrac{4\pi}{9} \left| \mathbf{d} \right|$ and~$C_2 = \tfrac{2\pi}{75} \sum_{ij} Q_{ij}^2$ (normalized to~$C_0 = 4\pi$ i.e.\ to~$\Phi_0=1$)
can also be computed as $C_\ell = \frac{1}{2\ell+1}\sum_{m=-\ell}^{+\ell} a_{\ell m}^2$.
\begin{table}
    \centering
    \begin{tabular}{c|ccc}
        energies (Auger) & $[8.57~\EeV, 16~\EeV)$ & $[16~\EeV, 32~\EeV)$ & $[32~\EeV, +\infty)$ \\
        energies (TA) & $[10~\EeV, 19.47~\EeV)$ & $[19.47~\EeV, 40.8~\EeV)$ & $[40.8~\EeV, +\infty)$ \\
        \hline
$d_x~[\%]$ & $-0.7\pm1.1\pm0.0$ & $+1.6\pm2.0\pm0.0$ & $-5.3\pm3.9\pm0.1$ \\
$d_y~[\%]$ & $+4.8\pm1.1\pm0.0$ & $+3.9\pm1.9\pm0.1$ & $+9.7\pm3.7\pm0.0$ \\
$d_z~[\%]$ & $-3.3\pm1.4\pm1.3$ & $-6.0\pm2.4\pm1.3$ & $+3.4\pm4.7\pm3.6$ \\
\hline
$Q_{xx}-Q_{yy}~[\%]$ & $-5.1\pm4.8\pm0.0$ & $+13.6\pm8.3\pm0.0\phantom{0}$ & $+43\pm16\pm0\phantom{0}$ \\
$Q_{xz}~[\%]$ & $-3.9\pm2.9\pm0.1$ & $+5.4\pm5.1\pm0.0$ & $+5\pm11\pm0$ \\
$Q_{yz}~[\%]$ & $-4.9\pm2.9\pm0.0$ & $-9.6\pm5.0\pm0.1$ & $+11.9\pm9.8\pm0.2\phantom{0}$ \\
$Q_{zz}~[\%]$ & $+0.5\pm3.3\pm1.7$ & $+5.2\pm5.8\pm1.7$ & $+20\pm11\pm5\phantom{0}$ \\
$Q_{xy}~[\%]$ & $+2.2\pm2.4\pm0.0$ & $+0.2\pm4.2\pm0.1$ & $+4.5\pm8.1\pm0.1$ \\
\hline
$C_1~[10^{-3}]$ & $4.8\pm2.0\pm1.2$ & $7.6\pm4.6\pm2.2$ & $19\pm12\pm4$\\
$C_2~[10^{-3}]$ & $0.85\pm0.66\pm0.02$ & $3.1\pm2.2\pm0.2$ & $15.5\pm8.9\pm2.4$\\
\hline
    \end{tabular}
    \caption{The dipole and quadrupole moments estimated from our data.  The first uncertainty is statistical, the second is due to that on the energy cross-calibration.
    The statistical uncertainties are uncorrelated ($-0.1 < \rho < 0.1$) except
    $\rho(d_x, Q_{xz}) = \rho(d_y, Q_{yz}) = 0.45$ and~$\rho(d_z, Q_{zz}) = 0.53$.}
    \label{tab:moments}
\end{table}
The results in the three energy bins defined above are shown in \autoref{tab:moments}.  
\begin{figure}
    \centering
    \includegraphics{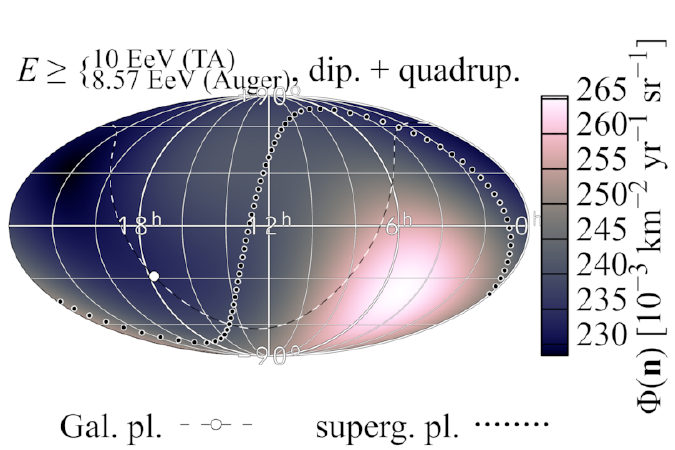}
    \hfil
    \includegraphics{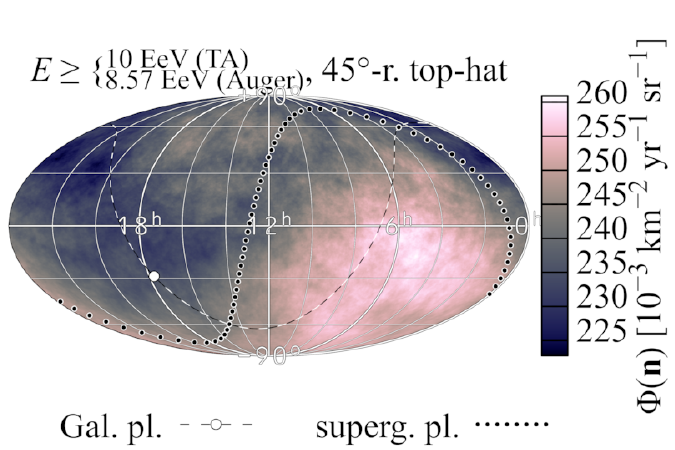}
    \includegraphics{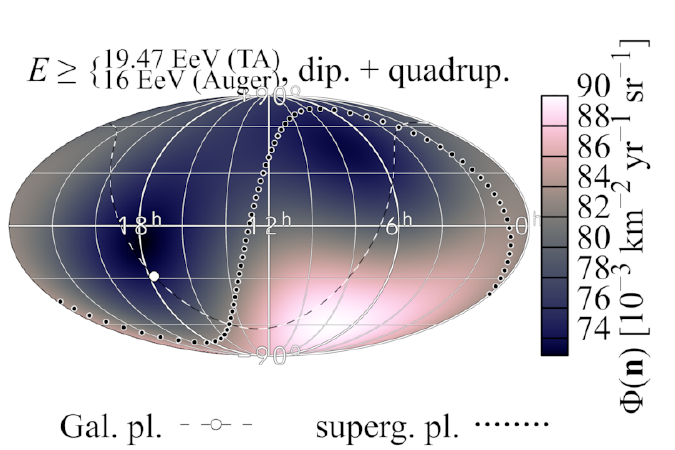}
    \hfil
    \includegraphics{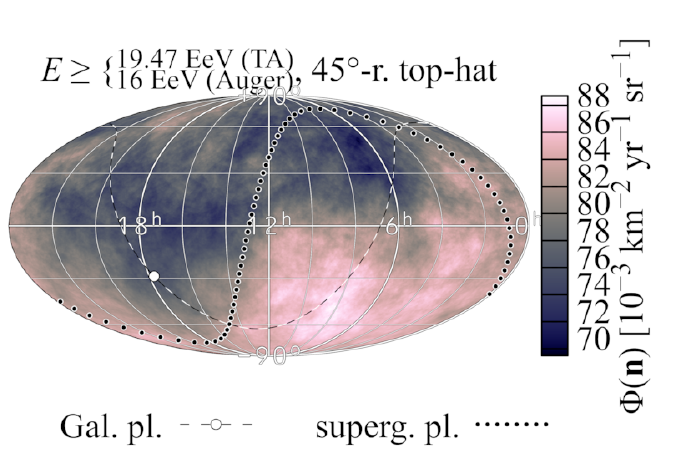}
    \includegraphics{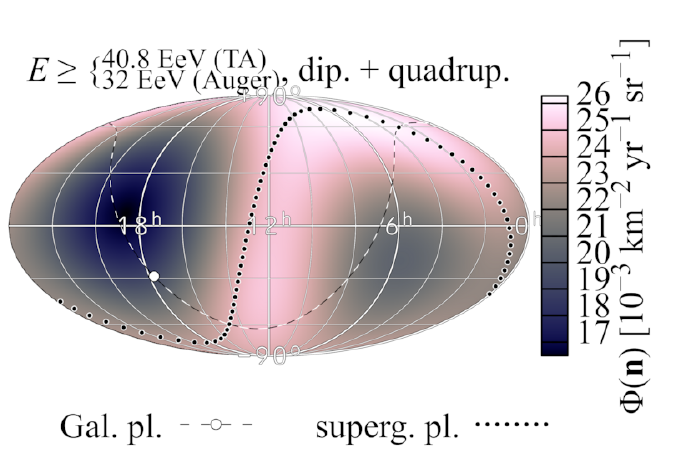}
    \hfil
    \includegraphics{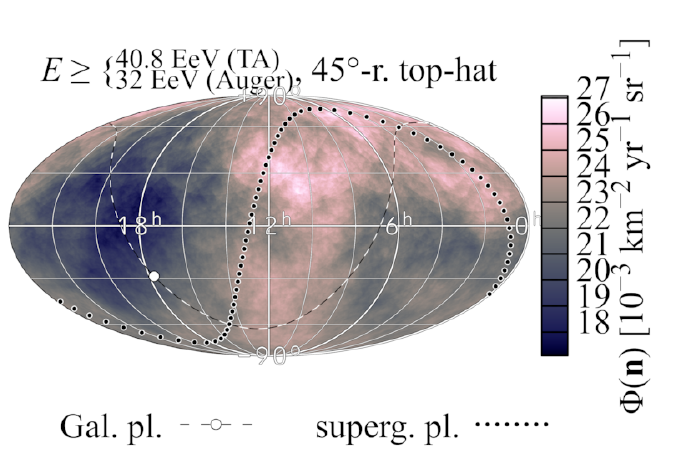}
    \caption{Left: Spherical harmonic expansion of the flux inferred from our data up to $\ell=2$ (dipole and quadrupole) in the three energy bins.  Right:  Measured flux averaged over $45\degree$-radius top-hat windows. }
    \label{fig:multipoles}
\end{figure}
The spherical harmonic expansion of the UHECR flux in these bins, truncated at $\ell \leq 2$, is shown in \autoref{fig:multipoles} in comparison with the flux estimate from the combined data in the corresponding bin, smoothed with the window size of $45^\circ$. 

\section{Discussion}

Using full-sky data, we have estimated the dipole and quadrupole moments of the UHECR flux distribution without any assumptions about higher-order multipoles.  The results are compatible with Auger-only ones assuming $\ell_{\max}=2$ \cite{bib:AugerLSA2021}, but with uncertainties on~$d_z$ and~$Q_{zz}$ about twice as small, as well as slightly smaller uncertainties on the other dipole and quadrupole components.
None of the moments shown in \autoref{tab:moments} are significant with $> 3\sigma$ pre-trial significance, except the dipole along the $y$~axis in the lowest energy bin already reported in Ref.~\cite{bib:AugerScience2017}.  

\newcommand{\etal}{et~al.}
\newcommand{\journal}[5]{\href{https://doi.org/#5}{\textit{#1}\ \textbf{#2} (#3)\ #4}}
\newcommand{\arXiv}[1]{\href{https://arxiv.org/abs/#1}{\nolinkurl{#1}}}

\clearpage
\section*{The Pierre Auger Collaboration}
\small

\begin{wrapfigure}[8]{l}{0.11\linewidth}
\vspace{-5mm}
\includegraphics[width=0.98\linewidth]{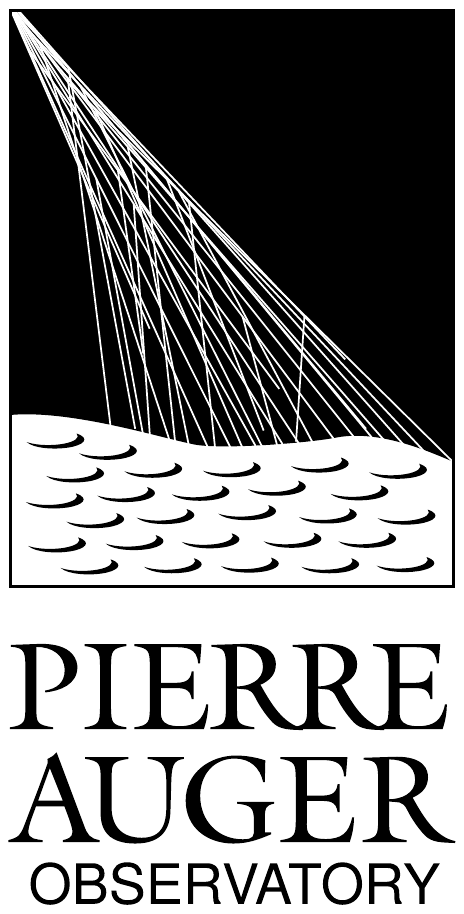}
\end{wrapfigure}
\begin{sloppypar}\noindent
\input{latex_authorlist_authors}
\end{sloppypar}

\begin{center}
\rule{0.1\columnwidth}{0.5pt}
\raisebox{-0.4ex}{\scriptsize$\bullet$}
\rule{0.1\columnwidth}{0.5pt}
\end{center}

\vspace{-1ex}
\footnotesize
\input{latex_authorlist_institutions}

\section*{The Telescope Array Collaboration}
\small

\begin{wrapfigure}[7]{l}{0.2\linewidth}
\vspace{-5mm}
\includegraphics[width=1.2\linewidth]{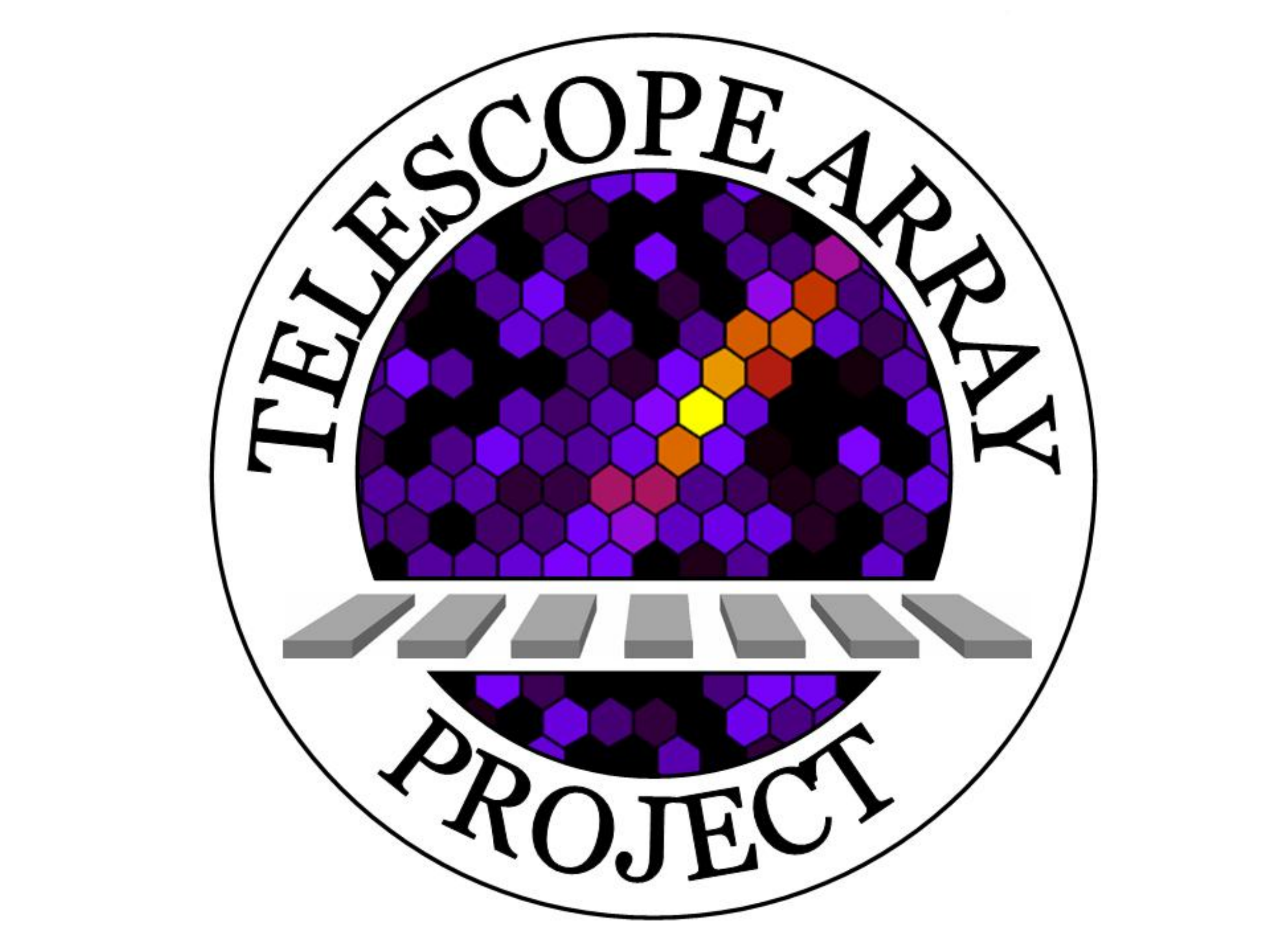}
\end{wrapfigure}
\begin{sloppypar}\noindent
\input{authors-ta}
\end{sloppypar}

\begin{center}
\rule{0.1\columnwidth}{0.5pt}
\raisebox{-0.4ex}{\scriptsize$\bullet$}
\rule{0.1\columnwidth}{0.5pt}
\end{center}

\vspace{-1ex}
\footnotesize
\input{institutions-ta}

\end{document}

%% file: latex_authorlist_authors.tex
P.~Abreu$^{72}$,
M.~Aglietta$^{54,52}$,
J.M.~Albury$^{13}$,
I.~Allekotte$^{1}$,
A.~Almela$^{8,12}$,
J.~Alvarez-Mu\~niz$^{79}$,
R.~Alves Batista$^{80}$,
G.A.~Anastasi$^{63,52}$,
L.~Anchordoqui$^{87}$,
B.~Andrada$^{8}$
% ...

%% file: latex_authorlist_institutions.tex

\begin{description}[labelsep=0.2em,align=right,labelwidth=0.7em,labelindent=0em,leftmargin=2em,noitemsep]
\item[$^{1}$] Centro At\'omico Bariloche and Instituto Balseiro (CNEA-UNCuyo-CONICET), San Carlos de Bariloche, Argentina
\item[$^{2}$] Centro de Investigaciones en L\'aseres y Aplicaciones, CITEDEF and CONICET, Villa Martelli, Argentina
\item[$^{3}$] Departamento de F\'\i{}sica and Departamento de Ciencias de la Atm\'osfera y los Oc\'eanos, FCEyN, Universidad de Buenos Aires and CONICET, Buenos Aires, Argentina
\item[$^{4}$] IFLP, Universidad Nacional de La Plata and CONICET, La Plata, Argentina
\item[$^{5}$] Instituto de Astronom\'\i{}a y F\'\i{}sica del Espacio (IAFE, CONICET-UBA), Buenos Aires, Argentina
\item[$^{6}$] Instituto de F\'\i{}sica de Rosario (IFIR) -- CONICET/U.N.R.\ and Facultad de Ciencias Bioqu\'\i{}micas y Farmac\'euticas U.N.R., Rosario, Argentina
\item[$^{7}$] Instituto de Tecnolog\'\i{}as en Detecci\'on y Astropart\'\i{}culas (CNEA, CONICET, UNSAM), and Universidad Tecnol\'ogica Nacional -- Facultad Regional Mendoza (CONICET/CNEA), Mendoza, Argentina
\item[$^{8}$] Instituto de Tecnolog\'\i{}as en Detecci\'on y Astropart\'\i{}culas (CNEA, CONICET, UNSAM), Buenos Aires, Argentina
\item[$^{9}$] International Center of Advanced Studies and Instituto de Ciencias F\'\i{}sicas, ECyT-UNSAM and CONICET, Campus Miguelete -- San Mart\'\i{}n, Buenos Aires, Argentina
\item[$^{10}$] Observatorio Pierre Auger, Malarg\"ue, Argentina
\item[$^{11}$] Observatorio Pierre Auger and Comisi\'on Nacional de Energ\'\i{}a At\'omica, Malarg\"ue, Argentina
\item[$^{12}$] Universidad Tecnol\'ogica Nacional -- Facultad Regional Buenos Aires, Buenos Aires, Argentina
\item[$^{13}$] University of Adelaide, Adelaide, S.A., Australia
\item[$^{14}$] Universit\'e Libre de Bruxelles (ULB), Brussels, Belgium
\item[$^{15}$] Vrije Universiteit Brussels, Brussels, Belgium
\item[$^{16}$] Centro Brasileiro de Pesquisas Fisicas, Rio de Janeiro, RJ, Brazil
\item[$^{17}$] Centro Federal de Educa\c{c}\~ao Tecnol\'ogica Celso Suckow da Fonseca, Nova Friburgo, Brazil
\item[$^{18}$] Instituto Federal de Educa\c{c}\~ao, Ci\^encia e Tecnologia do Rio de Janeiro (IFRJ), Brazil
\item[$^{19}$] Universidade de S\~ao Paulo, Escola de Engenharia de Lorena, Lorena, SP, Brazil
\item[$^{20}$] Universidade de S\~ao Paulo, Instituto de F\'\i{}sica de S\~ao Carlos, S\~ao Carlos, SP, Brazil
\item[$^{21}$] Universidade de S\~ao Paulo, Instituto de F\'\i{}sica, S\~ao Paulo, SP, Brazil
\item[$^{22}$] Universidade Estadual de Campinas, IFGW, Campinas, SP, Brazil
\item[$^{23}$] Universidade Estadual de Feira de Santana, Feira de Santana, Brazil
\item[$^{24}$] Universidade Federal do ABC, Santo Andr\'e, SP, Brazil
\item[$^{25}$] Universidade Federal do Paran\'a, Setor Palotina, Palotina, Brazil
\item[$^{26}$] Universidade Federal do Rio de Janeiro, Instituto de F\'\i{}sica, Rio de Janeiro, RJ, Brazil
\item[$^{27}$] Universidade Federal do Rio de Janeiro (UFRJ), Observat\'orio do Valongo, Rio de Janeiro, RJ, Brazil
\item[$^{28}$] Universidade Federal Fluminense, EEIMVR, Volta Redonda, RJ, Brazil
\item[$^{29}$] Universidad de Medell\'\i{}n, Medell\'\i{}n, Colombia
\item[$^{30}$] Universidad Industrial de Santander, Bucaramanga, Colombia
\item[$^{31}$] Charles University, Faculty of Mathematics and Physics, Institute of Particle and Nuclear Physics, Prague, Czech Republic
\item[$^{32}$] Institute of Physics of the Czech Academy of Sciences, Prague, Czech Republic
\item[$^{33}$] Palacky University, RCPTM, Olomouc, Czech Republic
\item[$^{34}$] CNRS/IN2P3, IJCLab, Universit\'e Paris-Saclay, Orsay, France
\item[$^{35}$] Laboratoire de Physique Nucl\'eaire et de Hautes Energies (LPNHE), Sorbonne Universit\'e, Universit\'e de Paris, CNRS-IN2P3, Paris, France
\item[$^{36}$] Univ.\ Grenoble Alpes, CNRS, Grenoble Institute of Engineering Univ.\ Grenoble Alpes, LPSC-IN2P3, 38000 Grenoble, France
\item[$^{37}$] Universit\'e Paris-Saclay, CNRS/IN2P3, IJCLab, Orsay, France
\item[$^{38}$] Bergische Universit\"at Wuppertal, Department of Physics, Wuppertal, Germany
\item[$^{39}$] Karlsruhe Institute of Technology (KIT), Institute for Experimental Particle Physics, Karlsruhe, Germany
\item[$^{40}$] Karlsruhe Institute of Technology (KIT), Institut f\"ur Prozessdatenverarbeitung und Elektronik, Karlsruhe, Germany
\item[$^{41}$] Karlsruhe Institute of Technology (KIT), Institute for Astroparticle Physics, Karlsruhe, Germany
\item[$^{42}$] RWTH Aachen University, III.\ Physikalisches Institut A, Aachen, Germany
\item[$^{43}$] Universit\"at Hamburg, II.\ Institut f\"ur Theoretische Physik, Hamburg, Germany
\item[$^{44}$] Universit\"at Siegen, Department Physik -- Experimentelle Teilchenphysik, Siegen, Germany
\item[$^{45}$] Gran Sasso Science Institute, L'Aquila, Italy
\item[$^{46}$] INFN Laboratori Nazionali del Gran Sasso, Assergi (L'Aquila), Italy
\item[$^{47}$] INFN, Sezione di Catania, Catania, Italy
\item[$^{48}$] INFN, Sezione di Lecce, Lecce, Italy
\item[$^{49}$] INFN, Sezione di Milano, Milano, Italy
\item[$^{50}$] INFN, Sezione di Napoli, Napoli, Italy
\item[$^{51}$] INFN, Sezione di Roma ``Tor Vergata'', Roma, Italy
\item[$^{52}$] INFN, Sezione di Torino, Torino, Italy
\item[$^{53}$] Istituto di Astrofisica Spaziale e Fisica Cosmica di Palermo (INAF), Palermo, Italy
\item[$^{54}$] Osservatorio Astrofisico di Torino (INAF), Torino, Italy
\item[$^{55}$] Politecnico di Milano, Dipartimento di Scienze e Tecnologie Aerospaziali , Milano, Italy
\item[$^{56}$] Universit\`a del Salento, Dipartimento di Matematica e Fisica ``E.\ De Giorgi'', Lecce, Italy
\item[$^{57}$] Universit\`a dell'Aquila, Dipartimento di Scienze Fisiche e Chimiche, L'Aquila, Italy
\item[$^{58}$] Universit\`a di Catania, Dipartimento di Fisica e Astronomia, Catania, Italy
\item[$^{59}$] Universit\`a di Milano, Dipartimento di Fisica, Milano, Italy
\item[$^{60}$] Universit\`a di Napoli ``Federico II'', Dipartimento di Fisica ``Ettore Pancini'', Napoli, Italy
\item[$^{61}$] Universit\`a di Palermo, Dipartimento di Fisica e Chimica ''E.\ Segr\`e'', Palermo, Italy
\item[$^{62}$] Universit\`a di Roma ``Tor Vergata'', Dipartimento di Fisica, Roma, Italy
\item[$^{63}$] Universit\`a Torino, Dipartimento di Fisica, Torino, Italy
\item[$^{64}$] Benem\'erita Universidad Aut\'onoma de Puebla, Puebla, M\'exico
\item[$^{65}$] Unidad Profesional Interdisciplinaria en Ingenier\'\i{}a y Tecnolog\'\i{}as Avanzadas del Instituto Polit\'ecnico Nacional (UPIITA-IPN), M\'exico, D.F., M\'exico
\item[$^{66}$] Universidad Aut\'onoma de Chiapas, Tuxtla Guti\'errez, Chiapas, M\'exico
\item[$^{67}$] Universidad Michoacana de San Nicol\'as de Hidalgo, Morelia, Michoac\'an, M\'exico
\item[$^{68}$] Universidad Nacional Aut\'onoma de M\'exico, M\'exico, D.F., M\'exico
\item[$^{69}$] Universidad Nacional de San Agustin de Arequipa, Facultad de Ciencias Naturales y Formales, Arequipa, Peru
\item[$^{70}$] Institute of Nuclear Physics PAN, Krakow, Poland
\item[$^{71}$] University of \L{}\'od\'z, Faculty of High-Energy Astrophysics,\L{}\'od\'z, Poland
\item[$^{72}$] Laborat\'orio de Instrumenta\c{c}\~ao e F\'\i{}sica Experimental de Part\'\i{}culas -- LIP and Instituto Superior T\'ecnico -- IST, Universidade de Lisboa -- UL, Lisboa, Portugal
\item[$^{73}$] ``Horia Hulubei'' National Institute for Physics and Nuclear Engineering, Bucharest-Magurele, Romania
\item[$^{74}$] Institute of Space Science, Bucharest-Magurele, Romania
\item[$^{75}$] University Politehnica of Bucharest, Bucharest, Romania
\item[$^{76}$] Center for Astrophysics and Cosmology (CAC), University of Nova Gorica, Nova Gorica, Slovenia
\item[$^{77}$] Experimental Particle Physics Department, J.\ Stefan Institute, Ljubljana, Slovenia
\item[$^{78}$] Universidad de Granada and C.A.F.P.E., Granada, Spain
\item[$^{79}$] Instituto Galego de F\'\i{}sica de Altas Enerx\'\i{}as (IGFAE), Universidade de Santiago de Compostela, Santiago de Compostela, Spain
\item[$^{80}$] IMAPP, Radboud University Nijmegen, Nijmegen, The Netherlands
\item[$^{81}$] Nationaal Instituut voor Kernfysica en Hoge Energie Fysica (NIKHEF), Science Park, Amsterdam, The Netherlands
\item[$^{82}$] Stichting Astronomisch Onderzoek in Nederland (ASTRON), Dwingeloo, The Netherlands
\item[$^{83}$] Universiteit van Amsterdam, Faculty of Science, Amsterdam, The Netherlands
\item[$^{84}$] University of Groningen, Kapteyn Astronomical Institute, Groningen, The Netherlands
\item[$^{85}$] Case Western Reserve University, Cleveland, OH, USA
\item[$^{86}$] Colorado School of Mines, Golden, CO, USA
\item[$^{87}$] Department of Physics and Astronomy, Lehman College, City University of New York, Bronx, NY, USA
\item[$^{88}$] Louisiana State University, Baton Rouge, LA, USA
\item[$^{89}$] Michigan Technological University, Houghton, MI, USA
\item[$^{90}$] New York University, New York, NY, USA
\item[$^{91}$] Pennsylvania State University, University Park, PA, USA
\item[$^{92}$] University of Chicago, Enrico Fermi Institute, Chicago, IL, USA
\item[$^{93}$] University of Delaware, Department of Physics and Astronomy, Bartol Research Institute, Newark, DE, USA
\item[$^{94}$] University of Wisconsin-Madison, Department of Physics and WIPAC, Madison, WI, USA
\item[] -----
\item[$^{a}$] Fermi National Accelerator Laboratory, Fermilab, Batavia, IL, USA
\item[$^{b}$] Max-Planck-Institut f\"ur Radioastronomie, Bonn, Germany
\item[$^{c}$] School of Physics and Astronomy, University of Leeds, Leeds, United Kingdom
\item[$^{d}$] Colorado State University, Fort Collins, CO, USA
\item[$^{e}$] now at Hakubi Center for Advanced Research and Graduate School of Science, Kyoto University, Kyoto, Japan
\item[$^{f}$] also at University of Bucharest, Physics Department, Bucharest, Romania
\end{description}

%% file: authors-ta.tex
R.U.~Abbasi$^{1,2}$,
T.~Abu-Zayyad$^{1,2}$,
M.~Allen$^{2}$,
Y.~Arai$^{3}$,
R.~Arimura$^{3}$,
E.~Barcikowski$^{2}$,
J.W.~Belz$^{2}$,
D.R.~Bergman$^{2}$,
S.A.~Blake$^{2}$,
I.~Buckland$^{2}$
% ...

%% file: institutions-ta.tex
{\footnotesize\it
$^{1}$ Department of Physics, Loyola University Chicago, Chicago, Illinois, USA \\
$^{2}$ High Energy Astrophysics Institute and Department of Physics and Astronomy, University of Utah, Salt Lake City, Utah, USA \\
$^{3}$ Graduate School of Science, Osaka City University, Osaka, Osaka, Japan \\
$^{4}$ Department of Physics and The Research Institute of Natural Science, Hanyang University, Seongdong-gu, Seoul, Korea \\
$^{5}$ Department of Physics, Tokyo University of Science, Noda, Chiba, Japan \\
$^{6}$ Institute for Cosmic Ray Research, University of Tokyo, Kashiwa, Chiba, Japan \\
$^{7}$ The Hakubi Center for Advanced Research and Graduate School of Science, Kyoto University, Kitashirakawa-Oiwakecho, Sakyo-ku, Kyoto, Japan \\
$^{8}$ Information Engineering Graduate School of Science and Technology, Shinshu University, Nagano, Nagano, Japan \\
$^{9}$ Faculty of Engineering, Kanagawa University, Yokohama, Kanagawa, Japan \\
$^{10}$ Interdisciplinary Graduate School of Medicine and Engineering, University of Yamanashi, Kofu, Yamanashi, Japan \\
$^{11}$ Academic Assembly School of Science and Technology Institute of Engineering, Shinshu University, Nagano, Nagano, Japan \\
$^{12}$ The Graduate School of Science and Engineering, Saitama University, Saitama, Saitama, Japan \\
$^{13}$ Astrophysical Big Bang Laboratory, RIKEN, Wako, Saitama, Japan \\
$^{14}$ Department of Physics, SungKyunKwan University, Jang-an-gu, Suwon, Korea \\
$^{15}$ Department of Physics, Tokyo City University, Setagaya-ku, Tokyo, Japan \\
$^{16}$ Institute for Nuclear Research of the Russian Academy of Sciences, Moscow, Russia \\
$^{17}$ Faculty of Systems Engineering and Science, Shibaura Institute of Technology, Minato-ku, Tokyo, Japan \\
$^{18}$ Department of Engineering Science, Faculty of Engineering, Osaka Electro-Communication University, Neyagawa-shi, Osaka, Japan \\
$^{19}$ Department of Physics, Chiba University, Chiba, Chiba, Japan \\
$^{20}$ Service de Physique Théorique, Université Libre de Bruxelles, Brussels, Belgium \\
$^{21}$ Department of Physics, Yonsei University, Seodaemun-gu, Seoul, Korea \\
$^{22}$ Center for Astrophysics and Cosmology, University of Nova Gorica, Nova Gorica, Slovenia \\
$^{23}$ Faculty of Science, Kochi University, Kochi, Kochi, Japan \\
$^{24}$ Nambu Yoichiro Institute of Theoretical and Experimental Physics, Osaka City University, Osaka, Osaka, Japan \\
$^{25}$ Department of Physical Sciences, Ritsumeikan University, Kusatsu, Shiga, Japan \\
$^{26}$ Quantum ICT Advanced Development Center, National Institute for Information and Communications Technology, Koganei, Tokyo, Japan \\
$^{27}$ Sternberg Astronomical Institute, Moscow M.V. Lomonosov State University, Moscow, Russia \\
$^{28}$ Department of Physics, School of Natural Sciences, Ulsan National Institute of Science and Technology, UNIST-gil, Ulsan, Korea \\
$^{29}$ Earthquake Research Institute, University of Tokyo, Bunkyo-ku, Tokyo, Japan \\
$^{30}$ Graduate School of Information Sciences, Hiroshima City University, Hiroshima, Hiroshima, Japan \\
$^{31}$ Institute of Particle and Nuclear Studies, KEK, Tsukuba, Ibaraki, Japan \\
$^{32}$ Graduate School of Science and Engineering, Tokyo Institute of Technology, Meguro, Tokyo, Japan \\
$^{33}$ Department of Research Planning and Promotion, Quantum Medical Science Directorate, National Institutes for Quantum and Radiological Science and Technology, Chiba, Chiba, Japan \\
$^{34}$ CEICO, Institute of Physics, Czech Academy of Sciences, Prague, Czech Republic \\
$^{35}$ Department of Physics and Institute for the Early Universe, Ewha Womans University, Seodaaemun-gu, Seoul, Korea
}

%% file: skeleton.bbl
\begin{thebibliography}{99}
\footnotesize\raggedright\setlength{\itemsep}{0pt}
\bibitem{bib:AugerScience2017} A.~Aab \etal\ [Pierre Auger collab.], \journal{Science}{357}{2017}{1266}{10.1126/science.aan4338} [\arXiv{1709.07321}].
\bibitem{Abbasi:2020ohd} R.U.~Abbasi \etal\ [Telescope Array collab.], \journal{Astrophys.\ J.\ Lett.}{898}{2020}{L28}{10.3847/2041-8213/aba0bc} [\arXiv{2007.00023}].
\bibitem{bib:ApJ2014} A.~Aab \etal\ [Pierre Auger and Telescope Array collabs.], \journal{Astrophys.\ J.}{794}{2014}{172}{10.1088/0004-637X/794/2/172} [\arXiv{1409.3128}].
\bibitem{bib:UHECR2016} A.~di~Matteo \etal\ [for the Pierre Auger and Telescope Array collabs.], \journal{JPS Conf.\ Proc.}{19}{2018}{011020}{10.7566/JPSCP.19.011020}.
\bibitem{bib:UHECR2018} J.~Biteau \etal\ [for the Pierre Auger and Telescope Array collabs.], \journal{EPJ Web Conf.}{210}{2019}{01005}{10.1051/epjconf/201921001005} [\arXiv{1905.04188}].
\bibitem{bib:Auger} A.~Aab \etal\ [Pierre Auger collab.], \journal{Nucl.\ Instrum.\ Meth.\ A}{798}{2015}{172}{10.1016/j.nima.2015.06.058} [\arXiv{1502.01323}].
\bibitem{bib:AugerLSA2021} R.~Menezes [for the Pierre Auger collab.], these proceedings.
\bibitem{bib:AugerPRD2020} A.~Aab \etal\ [Pierre Auger collab.], \journal{Phys.\ Rev.\ D}{102}{2020}{062005}{10.1103/PhysRevD.102.062005} [\arXiv{2008.06486}].
\bibitem{TAdetector} T.~Abu-Zayyad \etal\ [Telescope Array collab.], \journal{Nucl.\ Instrum.\ Meth.\ A}{689}{2012}{87}{10.1016/j.nima.2012.05.079} [\arXiv{1201.4964}].
\bibitem{bib:exposure} P.\ Sommers, \journal{Astropart.\ Phys.}{14}{2001}{271}{10.1016/S0927-6505(00)00130-4} [\arXiv{astro-ph/0004016}].
\bibitem{bib:spectrum2016} T.~Abu-Zayyad \etal\ [for the Pierre Auger and Telescope Array collabs.], \journal{JPS Conf.\ Proc.}{19}{2018}{011003}{10.7566/JPSCP.19.011003}.
\bibitem{bib:spectrum2018} T.~Abu-Zayyad \etal\ [for the Pierre Auger and Telescope Array collabs.], \journal{EPJ Web Conf.}{210}{2019}{01002}{10.1051/epjconf/201921001002} [\arXiv{1905.04188}].
\bibitem{bib:spectrumWG} Y.~Tsunesada [for the Pierre Auger and Telescope Array collabs.], these proceedings.

\end{thebibliography}
